# Co-development of significant elastic and reversible plastic deformation in nanowires

Peite Bao,[1] Yanbo Wang,[2] Xiangyuan Cui,[2,3] Qiang Gao,[4] Hongwei Liu,[3] Wai Kong Yeoh,[2,3] Hung-wei Yen,[2,3] Xiaozhou Liao,[2,*] Sichao Du,[1] H. Hoe Tan,[4] Chennupati Jagadish,[4] Jin Zou,[5] S. P. Ringer,[2,3] and Rongkun Zheng[1,*]

[1] *School of Physics, The University of Sydney, Sydney 2006, New South Wales, Australia, 2006, Australia*
[2] *School of Aerospace, Mechanical and Mechatronic Engineering, The University of Sydney, New South Wales, 2006, Australia*
[3] *Australian Centre for Microscopy and Microanalysis, The University of Sydney, New South Wale*
[3] *Department of Electronic Materials Engineering, Research School of Physics and Engineering, The Australian National University, Canberra, ACT 0200, Australia*
[5] *Materials Engineering and Centre for Microscopy and Microanalysis, The University of Queensland, QLD 4072, Australia*

**(Dated: March 12, 2013)**

When a material is subjected to an applied stress, the material will experience recoverable elastic deformation followed by permanent plastic deformation at the point when the applied stress exceeds the yield stress of the material[1]. Microscopically, the onset of the plasticity usually indicates the activation of dislocation motion, which is considered to be the primary mechanism of plastic deformation[2-4]. Once plastic deformation is initiated, further elastic deformation is negligible owing to the limited increase in the flow stress caused by work hardening. Here we present experimental evidence and quantitative analysis of simultaneous development of significant elastic deformation and dislocation-based plastic deformation in single crystal GaAs nanowires (NWs) under bending deformation up to a total strain of ~ 6%. The observation is in sharp contrast to the previous notions regarding the deformation modes. Most of the plastic deformation recovers spontaneously when the external stress is released, and therefore resembles an elastic deformation process.

It is known that the directional nature of ionic/covalent bonds in a ceramic crystal, *e.g.*, GaAs, makes dislocation slip difficult[5]. Therefore, in response to an applied stress, bulk crystalline ceramic materials fracture after a small elastic strain of ~ 0.1%, while dislocation mediated plastic strain is almost non-existent[6]. On the other hand, ceramic NWs are known to support large elastic strain primarily because of the constrained size and number of grains across the NW diameter due to the finite volume of a NW that leads to the dramatic reduction in the number of crystalline defects in individual NWs and effectively prevents the early initiation and propagation of crack in the NWs[7-9]. Semiconductor NWs with recoverable strain of ~ 3 – 10% has been reported and the strain has been described to be primarily elastic in nature[10-13]. However, recent research using computational simulations[14, 15] and nanoindentation experiments[16, 17] revealed that dislocation-based plasticity can also be reversible in some nanocrystalline materials. Therefore, it is necessary to conduct careful atomic-scale characterization of the deformation-recovery processes of NWs, as the stress that introduced large strain might be high enough to trigger significant plastic deformation. In this work, we develop a novel experimental method that enables atomic-scale characterization of nanostructures under any deformation level.

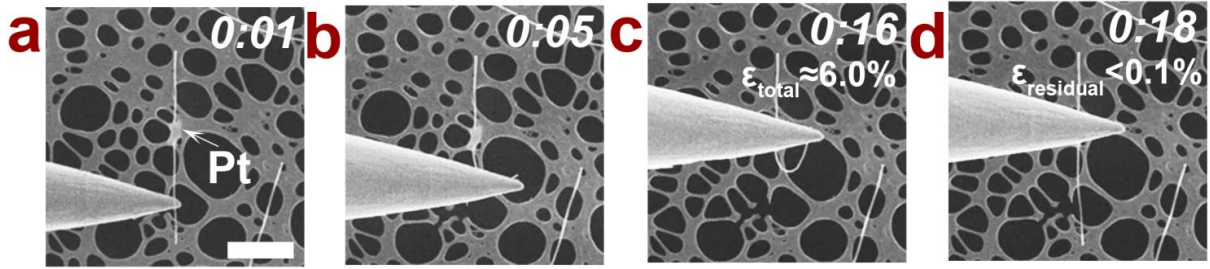

**Figure 1 | A typical deformation process of the GaAs NWs with a diameter of ~ 50 nm.** A series of SEM graphs extracted from Movie 1 in the Supplementary Information, labels on the top right corner indicate the time point when the graphs were extracted (in seconds). The largest strain $\varepsilon$ ~ 6% was reached in **c**, and the NW recovered to a straight shape with only very small plastic strain < 0.1% remained as in **d**. Scale bar: 2 µm.

Twin-free GaAs NWs were grown by using metalorganic chemical vapour deposition (MOCVD) via the vapor-liquid-solid mechanism using Au nanoparticles as catalyst. Trimethylgallium and AsH$_3$ were used as the Ga and As sources, respectively. A two-temperature procedure, which consists a 1 minute high-temperature (450ºC) nucleation step followed by a 30 minutes low-temperature (375ºC) growth step, was used and ensured samples were minimum tapered and of perfect crystalline quality[18]. NWs were vertically aligned on (111)B GaAs substrate and uniform in diameter of ~ 50 – 60 nm. NWs studied were ~ 4 – 6 µm in length. HRTEM investigation was performed on a JEOL 2200FS field-emission TEM operating at 200 kV and a JEOL 3000F field-emission TEM operating at 300 kV with a standard double-tilt mechanism. (111) d-spacing before, during and after deformation, are averaged values from three separate regions, each contains 15 – 20 lattices. Dislocation lines were identified via using inverse fast Fourier transformation (IFFT) of the HRTEM images. TEM sample preparation was done simply by scratching the growth substrate using copper grids with holey carbon supporting film. Deformation experiments were performed in a Zeiss Auriga dual-beam SEM equipped with a platinum gas injection system (GIS), and a Kleindiek Nanotechnik nanopositioning system connected with an electro-polished tungsten tip as the micro-manipulator inside of the SEM chamber. GIS enables to affix the NW to the carbon supporting film by depositing platinum patches onto selected areas. Controllable strain can then be applied to a single NW with the tungsten manipulator and maintained on the TEM sample with deposition. The releasing of the strain can be done by breaking either the deposition or the surrounded supporting film. Note that the SEM stage had been adjusted in order to find the largest curvature radius to obtain an accurate total strain $\varepsilon_{total}$.

A typical complete bending deformation process of a GaAs NW is shown in Figure 1, a series of scanning electron microscopy (SEM) images extracted from the movie recorded. A NW with a diameter of ~ 50 nm was affixed to a supporting carbon film on a copper grid by a small patch of platinum. A tungsten micro-manipulator was then used to progressively apply a torque to bend the NW in a controlled manner. The radius of the bending curvature, $R$, was measured in real-time and the total strain, $\varepsilon_{total}$, was calculated using the formula $\varepsilon_{total} = d/2R$, where $d$ is the NW diameter. As marked in Fig. 1c, the largest strain reached ~ 6% when the NW was bent almost 180°. When the external load was released, the NW almost returned to its initial straight status, as shown in Fig. 1d, with very little residual strain ($\varepsilon_{residual} < 0.1\%$). The residual strain is a proof that plastic deformation was initiated during the deformation process. It was also observed that GaAs NWs became so fragile after exposure to 6% strain that slight external turbulence, such as the gas flow from the platinum gas injection system, or movement of the micro-manipulator could trigger a fracture at the region which was most deformed.

The microstructural evolution of a NW during the deformation–recovery process is presented in Fig. 2. The NW with a diameter of ~ 56.4 nm was deformed to a strain of ~ 4.81%, representing the case of high strain deformation. High-resolution transmission electron microscopy (HRTEM) image in Fig. 2a shows that the strain contrast overlaps the lattice fringe. A selected-area electron diffraction (SAED) pattern along the [1 $\bar{1}$ 0] zone axis in the inset of Fig. 2a shows that diffraction spots were elongated to arcs due to the continuous bending of the NW. Careful examination of the crystal lattice reveals expansion at the outer-side and compression at the inner-side of the lattice, suggesting the existence of elastic deformation. Enlarged areas near the inner/outer NW edges, shown in Fig. 2b, were used to measure the d-spacing of the (111) planes along the NW axial direction, and line profiles extracted from Fig. 2b are displayed in Fig. 2c. The spacing for these (111) planes deviated substantially from the standard value of 3.26 Å before deformation. Multiple regions near the inner and outer edges at the most deformed areas were measured to obtain reliable average values for the inner and outer (111) plane d-spacing, $d_{inner}$ and $d_{outer}$, respectively. These values were then employed to estimate the elastic strain $\varepsilon_{elastic}$ caused by lattice distortion using equation $\varepsilon_{elastic} = (d_{outer} - d_{inner})/(d_{outer} + d_{inner})$ [12]. Interestingly, the calculated elastic strain $\varepsilon_{elastic}$ was ~ 3.06%, which is significantly smaller than the total strain $\varepsilon_{total}$ of 4.81%. It is expected that plastic deformation occurred and the plastic strain $\varepsilon_{plastic}$ accounts for the difference between $\varepsilon_{total}$ and $\varepsilon_{elastic}$, i.e., $\varepsilon_{plastic} = \varepsilon_{total} - \varepsilon_{elastic}$. $\varepsilon_{plastic}$ is ~ 1.75% for the NW with $\varepsilon_{total}$ of 4.81%.

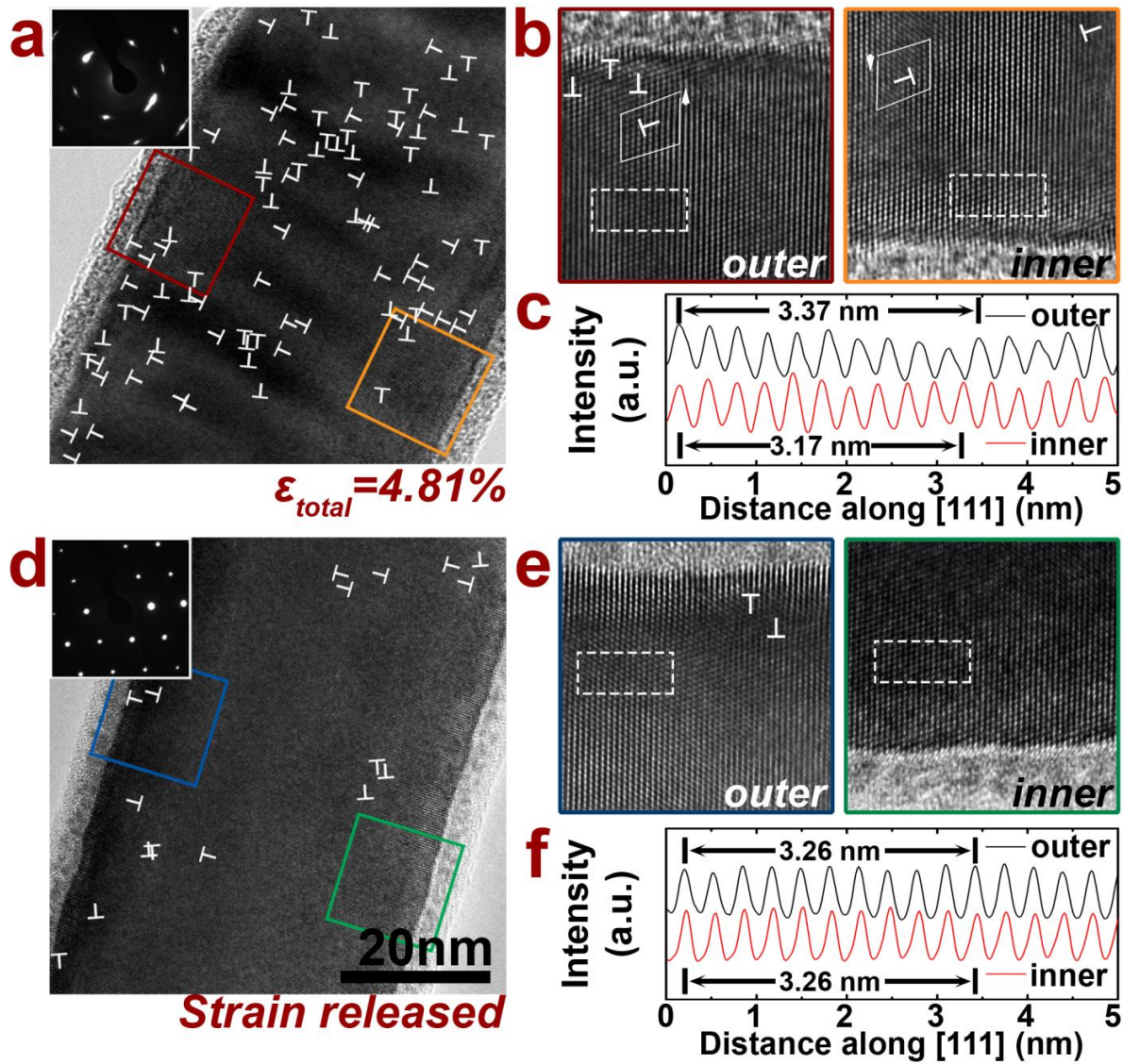

**Figure 2 | Microstructural evolution of GaAs NWs underwent a complete deformation/recovery process with different strain levels. a**, HRTEM image of a 4.81% deformed NW and the corresponding SAED pattern. A very high density of dislocations is seen. **b**, Enlarged from regions with corresponding colour frame from **a**, Burgers circuits were drawn to decide the dislocation type. **c**, Line intensity profiles from dashed rectangles in **b** showing the distorted (111) d-spacing under strain. **d**, The NW recovered spontaneously from **a** to straight shape, the dislocation density reduced dramatically and residual defects were mostly found near the edge of the NW. Inset SAED pattern is identical to that of the original samples. **e**, Enlarged regions from **d**. **f**, Line intensity profiles from dashed rectangles in **e** showing the recovery of the bond stretching/compressing. 'T' marks were used to show the dislocations lines in HRTEM images. In **c** and **f**, distances of ten lattices have been marked.

Indeed, dislocations were extensively observed in the deformed GaAs NW, as marked with "T" in Fig. 2a, which is a clear evidence of dislocation-based plastic deformation. Detailed Burgers circuit analysis (Fig. 2b) indicated that the dislocations were all 1/2<110> type perfect dislocations. Neither nano-twin nor extended dislocation/stacking fault was observed in the deformed NWs. It has been reported that nucleation and gliding of perfect dislocations is responsible for plastic deformation in bulk GaAs[19, 20]. In this work, plastic deformation in GaAs NWs was also formed through the generation and motion of perfect dislocations with 1/2<110> Burgers vectors. Dislocations in NWs or nanofibers are nucleated primarily from the surface once the applied stress surpasses a critical value, which is much smaller than the stress needed for dislocation self-multiplication[21].

Fig. 2d shows the microstructure of the NW after the stress was released, in which the dislocation density was dramatically reduced, leaving only a low density of dislocations close to the edge area of the NW, resulting in a residual strain $\varepsilon_{residue}$, typically < 0.1%.. The inset SAED pattern shows that the diffraction arcs transform back to diffraction spots. HRTEM images in Fig. 2e show the restoration to high crystal quality. Measurements of the (111) planar spacing, illustrated by the line profile in Fig. 2f, confirmed that the elastic deformation had fully recovered. Surprisingly, the results indicate that dislocation-based plastic deformation has almost completely recovered spontaneously. The reversibility of the plasticity resembles an elastic deformation process.

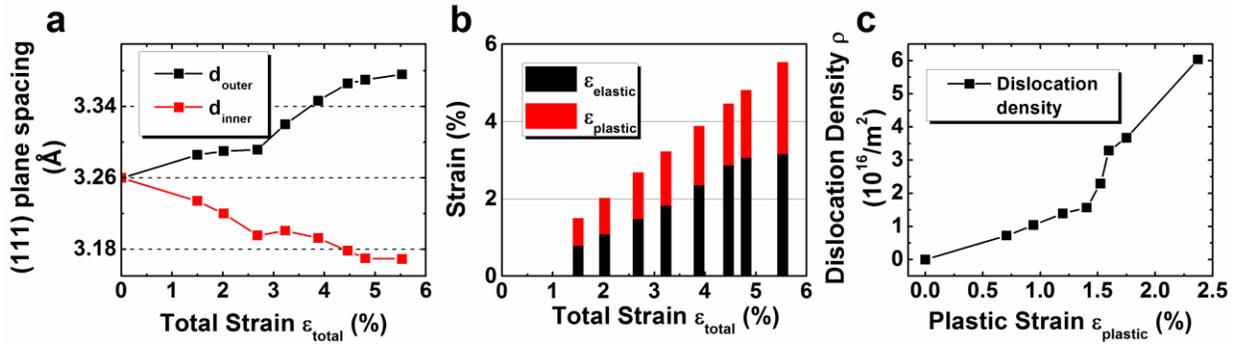

**Figure 3 | Statistics on GaAs nanowire microstructure within the recoverable deformation region. a**, Average (111) d-spacing values versus the total strain. **b**, Elastic and plastic strain components with respect to the total strain, illustrating that elastic deformation and plastic deformation occur simultaneously and proportions of each components stayed approximately unchanged. **c**, Dislocation density versus the plastic strain.

To investigate the behaviours and interaction of the elastic and plastic deformation, a number of NWs were deformed to different strain levels and the structural information was quantified, as shown in Fig. 3 (see Supplementary Information for detailed micrographs). Fig. 3a presents $d_{outer}$ and $d_{inner}$ as a function of $\varepsilon_{total}$. The proportions of elastic and plastic strain components as a function of total strain were deduced and presented in Fig. 3b. Both $\varepsilon_{elastic}$ and $\varepsilon_{plastic}$ increase with total strain, and the $\varepsilon_{elastic}/\varepsilon$ ratio stays approximately constant at ~ 60±5%. This result indicates the co-development of significant amount of elastic and plastic deformation, which is in sharp contrast to the conventional wisdom for bulk material that elastic deformation is negligible once plastic deformation is initiated. It is also in contradiction to the previous reports that NWs are free of dislocations within the elastic regime[22]. Dislocation densities, denoted by $\rho$, were calculated from the HRTEM images of the strained NWs, as plotted in Fig. 3c. The dislocation densities are on the order of $10^{16}$ m$^{-2}$, which are significantly higher than those in high-quality bulk GaAs single crystals ($10^5$ m$^{-2}$)[23]. From Fig. 3c, it is expected that when the plastic strain is over 2.5% or the total strain is larger than 6%, the dislocation density would exceed $10^{17}$ m$^{-2}$, which is too high for the NW to sustain and therefore fracture occurs[24].

The nanocrystalline structure and the bending deformation mode are believed responsible for the simultaneous development of significant elastic and plastic deformation from low strain until fracture. The NW surface provides abundant sources for dislocation nucleation due to the high surface to volume ratio in the NWs, making it possible to produce a very high density of dislocations during a deformation process. The nucleation of dislocations at the surface or grain boundaries usually requires much lower stress than the self-multiplication of dislocations by, *e.g.*, Frank-Read sources[25, 26] in nanocrystalline materials. For deformation with very large strain, dislocation multiplication could even be triggered. The magnitude of critical resolved shear stress for dislocation multiplication $\tau_{critical}$ is estimated to be ~ 3 GPa using the Kuhlmann-Wilsdorf law of similitude and $\tau_{critical} = K\mu b_p D^{-1}$, where, $K$ is proportionality factor that was decided from Ref 21, $\mu$ is the shear modulus ~ 3.29·$10^{11}$ dyn cm$^{-2}$, $b_p$ is the magnitude of Burgers vector (1/6<121>), and $D$ is the operating distance of a dislocation multiplication source that cannot be larger than the diameter of the NWs[23, 27]. In this work, $D$ was chosen to be the NW diameter $d$, which is ~ 50 nm, hence $\tau_{critical} = K\mu b_p d^{-1}$. On the other hand, elastic stress can be calculated with $\tau_{elastic} = E\varepsilon_{elastic}$, where the Young's modulus $E$ of the GaAs NWs with a diameter of ~ 50 nm was ~180 GPa[13]. Hence, when $\varepsilon_{elastic}$ reaches ~ 1.70%, dislocation multiplication can be triggered as the $\tau_{elastic}$ at this point reaches the $\tau_{critical}$ value for 50 nm GaAs NWs. In this work, the applied stress can be high enough to activate self-multiplication of dislocations after a certain amount of elastic deformation, although surface sources may still prevail.

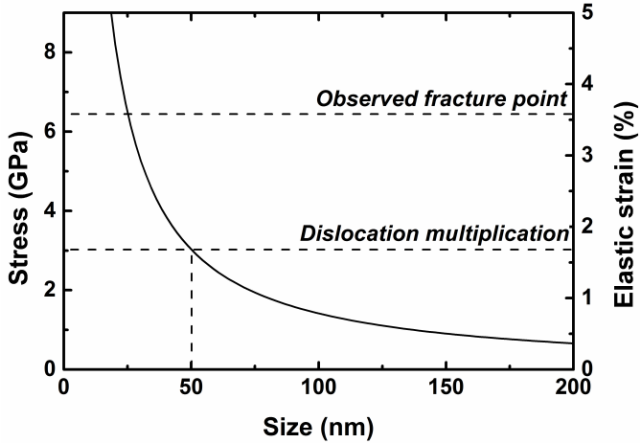

**Figure 4 | The relationship between the operating distance and critical resolved shear stress for GaAs, and the correspondence of the stress to the elastic strain.** The curve shows the relationship of $\tau_{critical} \propto D^{-1}$, and the strain values on the right axis is calculated from the stress values on the curve by $\tau_{elastic} = E\varepsilon_{elastic}$. $\varepsilon_{elastic}$ of ~ 1.70% corresponds to the stress required for dislocation multiplication for 50 nm GaAs NWs, and ~ 3.6% $\varepsilon_{elastic}$ refers to the fracture point observed from the experiments.

For bending deformation of a NW, the strain type changes from tension at one side to compression at the other side of the NW. The magnitude of strain gradually decreases from the edge to centre, so the stress minimizes somewhere around the geometric axial centre of the NW. Such a low stress area acts as a barrier to prevent dislocations from passing through the whole NW. Therefore, dislocations generated on the crystalline surface and by multiplication at regions with large strain build up rapidly in the NW, reaching an extremely high dislocation density on the order of $10^{16}$ m$^{-2}$. Consequently very significant work hardening occurs because of the repulsive force between dislocations with the same Burgers vectors and the tangling of dislocations that glide on different slip systems. This leads to a very rapid increase of the flow stress during the deformation, which is distinctive from the conventional case in coarse-grained materials where flow stress increase is limited during plastic deformation processes. The significant elevation in flow stress results in further increase of significant elastic strain after the initiation of plastic deformation.

When the applied stress is removed, the large internal stress is released through the recovery of elastic strain. The internal stress is opposite in direction and approximately equivalent in magnitude to the applied stress during the initial recovery process. This internal stress spontaneously activated and drove the dislocation motion backwards. It is known that surface or grain boundaries of nanomaterials acts as both the source and sink of dislocations[26, 28]. Hence, the reverse motion of the dislocations led to the reversal of the plasticity, which has been described by molecular dynamics simulations[14]. When most of the dislocations vanished at the surface, plastic strain is largely recovered, resembling the process of elastic deformation.

In summary, our study discovered that significant elastic and dislocation-based plastic strain occurs simultaneously during the bending deformation process of GaAs NWs, which is in sharp contrast to the textbook knowledge that elastic and plastic deformation are two different regions in materials deformation. This work indicates that the super recoverable deformation is not simply elastic[10-13] or reversible plastic[14-17] deformation in nature, but the coupling of both, and that caution is needed in evaluating the effect of elastic strain on the optoelectronic properties of semiconductors.

This work was supported in part by the Australian Research Council. The authors acknowledge the facilities and technical assistance from staff at the Australian Microscopy & Microanalysis Research Facility (AMMRF) at the University of Sydney. Growth facilities used in this work were supported by the Australian National Fabrication Facility (ANFF) at the Australian National University.

-----------------------------
*rongkun.zheng@sydney.edu.au;
xiazhou.liao@sydney.edu.au.